\documentclass[twocolumn,twocolappendix]{aastex6}

\usepackage{amsmath}

\begin{document}

\title{Bayesian estimation of thermonuclear reaction rates 

for deuterium+deuterium reactions}

\author{\'{A}. G\'{o}mez I\~{n}esta\altaffilmark{1}, C. Iliadis\altaffilmark{2,3} and A. Coc\altaffilmark{4}}
\email[]{alvarogomezinesta@gmail.com}
\email[]{iliadis@physics.unc.edu}
\altaffiliation
\altaffiltext{1}{}
\altaffiltext{1}{Department of Physics, Universitat Polit\`{e}cnica de Catalunya, Barcelona, 08930, Spain}
\altaffiltext{2}{Department of Physics \& Astronomy, University of North Carolina at Chapel Hill, Chapel Hill, NC 27599-3255, USA}
\altaffiltext{3}{Triangle Universities Nuclear Laboratory, Durham, NC 27708-0308, USA}
\altaffiltext{4}{Centre de Sciences Nucl\'{e}aires et de Sciences de la Mati\`{e}re (CSNSM), Univ. Paris-Sud, CNRS/IN2P3, Universit\'{e} Paris-Saclay, B\^{a}timent 104, F-91405 Orsay Campus, France}

\begin{abstract}
The study of d+d reactions is of major interest since their reaction rates affect the predicted abundances of D, $^3$He, and $^7$Li. In particular, recent measurements of primordial D/H ratios call for reduced uncertainties in the theoretical abundances predicted by big bang nucleosynthesis (BBN). 
Different authors have studied reactions involved in BBN by incorporating new experimental data and a careful treatment of systematic and probabilistic uncertainties. 
To analyze the experimental data, Coc et al. (2015) used results of \textit{ab initio} models for the theoretical calculation of the energy dependence of S-factors in conjunction with traditional statistical methods based on $\chi^2$ minimization. 
Bayesian methods have now spread to many scientific fields and provide numerous advantages in data analysis. Astrophysical S-factors and reaction rates using Bayesian statistics were calculated by Iliadis et al. (2016). Here we present a similar analysis for two d+d reactions, d(d,n)\textsuperscript{3}He and d(d,p)\textsuperscript{3}H, that has been translated into a total decrease of the predicted D/H value by 0.16\%.
\end{abstract}



\maketitle

\section{Introduction}
Big bang nucleosynthesis (BBN) is responsible for the formation of primordial \textsuperscript{2}H, \textsuperscript{3}He, \textsuperscript{4}He and \textsuperscript{7}Li. 
Considering that the primordial abundances of these isotopes span more than eight orders of magnitude, there is a fair agreement between BBN predictions and observations  (see \citet{cyb} for a recent review).
In recent years the uncertainties have been greatly reduced on both the primordial abundances deduced from observations, and on the parameters entering into the BBN model. 
For instance, observations of the anisotropies of the cosmic microwave background (CMB), e.g. by the Planck space mission \citep{planck2}, led to precise estimations of cosmological parameters. In
particular, the baryonic density of the Universe was measured with an uncertainty of less than 1\%: $\Omega_{\mathrm{b}}{\cdot}h^2$ = 0.02225$\pm$0.00016 \citep{planck2}. 
With this determination, the BBN model becomes parameter free and should be able to make accurate predictions.

However, it is now widely known (see \citet{Fie11} for a review) that there 
is a factor of three difference between the calculated \textsuperscript{7}Li/H ratio, by number of atoms \citep{cyb,coc}, and the corresponding primordial value deduced from observations  \citep{Sbo10}.
The primitive lithium abundance is deduced from observations of low metallicity stars in the halo of our Galaxy, where the lithium abundance is almost independent of metallicity, 
displaying a plateau both as a function of metallicity and effective temperature.
This puzzling discrepancy, known as the {\em lithium problem}, has not yet found a satisfactory solution \citep{NIC2016} and casts a shadow on the model.

The uncertainty on the \textsuperscript{4}He primordial abundance, which is deduced from the observation of metal--poor extragalactic H II regions, has been reduced by the inclusion 
of an additional atomic infrared line in the analysis \citep{Ave15}. For this isotope, BBN predictions agree well with observations, keeping in mind that these predictions rely on the n$\leftrightarrow$p 
weak reaction rates. One should note that these calculated rates incorporate various corrections that need to be assessed. The weak rates 
are also normalized to the experimental neutron lifetime whose recommended value, $\tau_{\rm{n}}$ = 880.3$\pm$1.1~s \citep{PDG}, has evolved in the last few years \citep{You14}.

Because of its low abundance, \textsuperscript{3}He, has not been observed outside of our Galaxy \citep{Ban02}. Since it is  both produced and destroyed in stars, its galactic chemical evolution is uncertain. 
It is, hence, presently of little use to constrain BBN. However, the next generation of 30+ m 
telescope facilities may allow to extract the $^3$He/$^4$He ratio from observations of extra-galactic metal poor HII regions \citep{Coo15}.

Deuterium's most primitive abundance is determined from the observation of few cosmological clouds at high redshift, on the line of sight of distant quasars. Up to a few years ago,
there was a significant scatter in observations that lead to an $\approx$8\% \citep{Oli12} uncertaininty on the primordial deuterium abundance. BBN prediction were, then, fully
compatible with observations. However, recent measurements of primordial D/H, based on observations of damped Lyman-$\alpha$ systems at high redshift, led to an uncertainty of 1.3\%, D/H = 
(2.547$\pm$0.033)$\times10^{-5}$ \citep{cooke}. This has to be compared to the most recent predictions of (2.45$\pm$0.05)$\times10^{-5}$ \citep{coc} and (2.58$\pm$0.04)$\times10^{-5}$ \citep{cyb}
that quote a 1.6--2.0\% uncertainty, but whose central values differ by 5\%. However, this difference almost vanishes if the same rates are used for the d(p,$\gamma)^3$He,
d(d,n)\textsuperscript{3}He and d(d,p)\textsuperscript{3}H nuclear reactions (Tsung-Han Yeh, priv. comm.). 
These small, but significant, differences between obeservations and predictions require further investigations that are currently underway, in particular, the re-evaluations
of reaction rates including the particle physics corrections to the weak rates,
the comparison between numerical methods used in the network calculations and the
comparison with other independent BBN codes and networks (e.g. \citet{cyb}).
This paper concerns one important contribution to this goal, but others are needed before
one is able to provide improved BBN predictions. This is why we will, here, only discuss
relative effects of these new rates.

An improved D/H predicion is also very important for the lithium problem since most proposed solutions lead to an unacceptable increase of the 
deuterium abundance \citep{Oli12,Kus14,coc}.   
Indeed, for the CMB deduced baryonic density, \textsuperscript{7}Li is produced, during primordial nucleosynthesis,
indirectly by $^3$He($\alpha,\gamma)^7$Be, where $^7$Be will decay much later to \textsuperscript{7}Li, while $^7$Be is destroyed by  
$^7$Be(n,p)$^7$Li(p,$\alpha)^4$He. The solutions to the lithium problem generally rely on an increased late time neutron abundance 
to boost $^7$Be destruction  through the $^7$Be(n,p)$^7$Li(p,$\alpha)^4$He channel. These extra neutrons, inevitably, also boost the deuterium  production through the  $^1$H(n,$\gamma)^2$H channel. 

Hence, it is very important that the uncertainties on D/H predictions be reduced, because ($i$) the observational uncertainties of the primordial D/H ratio are smaller than those predicted by simulations,
($ii$) differences appear between predictions using different prescriptions for the reaction rates, and ($iii$) deuterium provides strong constraints to solutions of the lithium problem.

The precision of these calculations is currently limited by our knowledge of certain key thermonuclear reaction rates. For example, a 10\% error in the  d(p,$\gamma)^3$He, d(d,n)\textsuperscript{3}He and 
d(d,p)\textsuperscript{3}H rates causes a 3.2\%, 5.4\% and 4.6\% uncertainty, respectively, in the predicted D/H ratio \citep{coc}.
The aim of our study is to reduce the uncertainties of BBN nucleosynthesis simulations as a continuation of our previous work that included the  d(p,$\gamma)^3$He rate \citep{iliadis}.
 Both d(d,n)\textsuperscript{3}He and d(d,p)\textsuperscript{3}H are non-resonant reactions, meaning that the S-factor, S(E), varies smoothly with energy. We apply a Bayesian analysis to the most recent experimental d+d S-factor data, and use the resulting improved S-factors to calculate the reaction rates. The theoretical model used for the S-factor \citep{arai} is assumed to accurately predict the energy dependence but not necessarily its absolute scale. The experimental data is used to scale this S-factor curve.
We carry out a multiparametric estimation. The model parameters are the scale factor for the theoretical S-factor (we will refer to it as ``overall scale factor" or ``scale factor") and a normalization factor for each data set accounting for systematic errors (we will refer to them as ``normalization factors"). Hence, there is a total of 6 parameters for each reaction, since there is an overall scale factor and 5 normalization factors, one per data set.
The Bayesian model provides a consistent description of all uncertainties involved (statistical and systematic), and yields the probability density for each parameter. Unlike traditional data analysis methods (e.g., \citealp{coc}), it does not involve \textit{ad hoc} assumptions or rely on Gaussian approximations for uncertainties. A more detailed explanation of this statistical analysis is given in Section \ref{strategy}. See \citet{iliadis} for further information on these Bayesian models.

\begin{figure*}[ht]
\begin{center}
			\centering
			\includegraphics[width=0.95\textwidth]{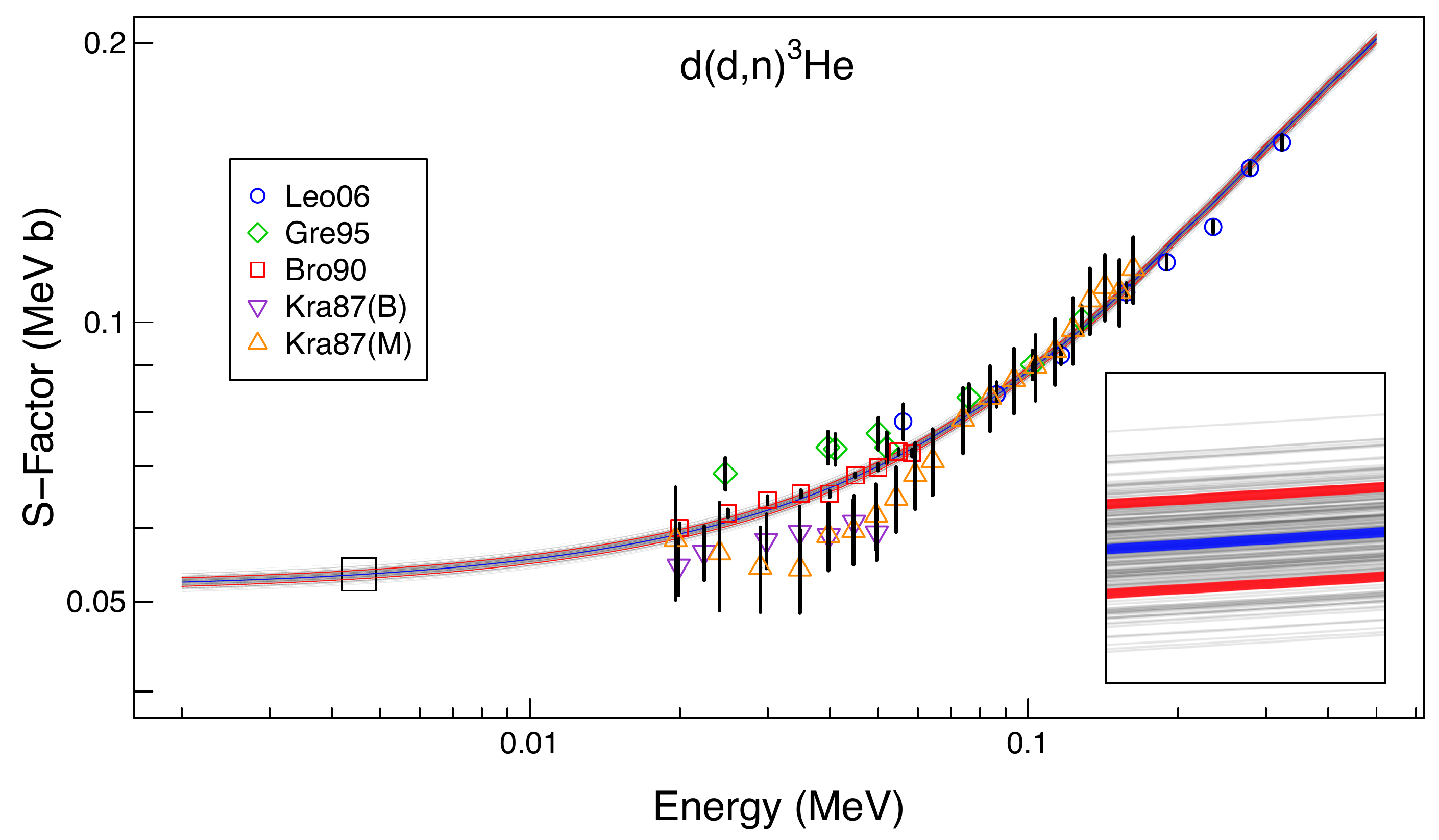}
			\caption{Astrophysical S-factor versus center-of-mass energy for the d(d,n)\textsuperscript{3}He reaction. The symbols show the data of \citet{leo} (circles), \citet{gre} (diamonds), \citet{bro} (squares), \citet{kra} (B) (down-pointing triangles) and \citet{kra} (M) (up-pointing triangles). The error bars (1$\sigma$) refer to statistical uncertainties only. Grey lines forming the shaded area correspond to credible S-factors that result from different sets of parameter samples (the inset shows a magnification for a clearer view of these lines). The blue line is the median (50th percentile) of all credible S-factors, and red lines correspond to the 16th and 84th percentiles. The credible lines are calculated from the theoretical S-factor of \citet{arai}, multiplied by a scale factor that is a parameter of the Bayesian model.}
			\label{fig:ResultsDdn}
\end{center}
\end{figure*}

\begin{figure*}[ht]
\begin{center}
			\centering
			\includegraphics[width=0.95\textwidth]{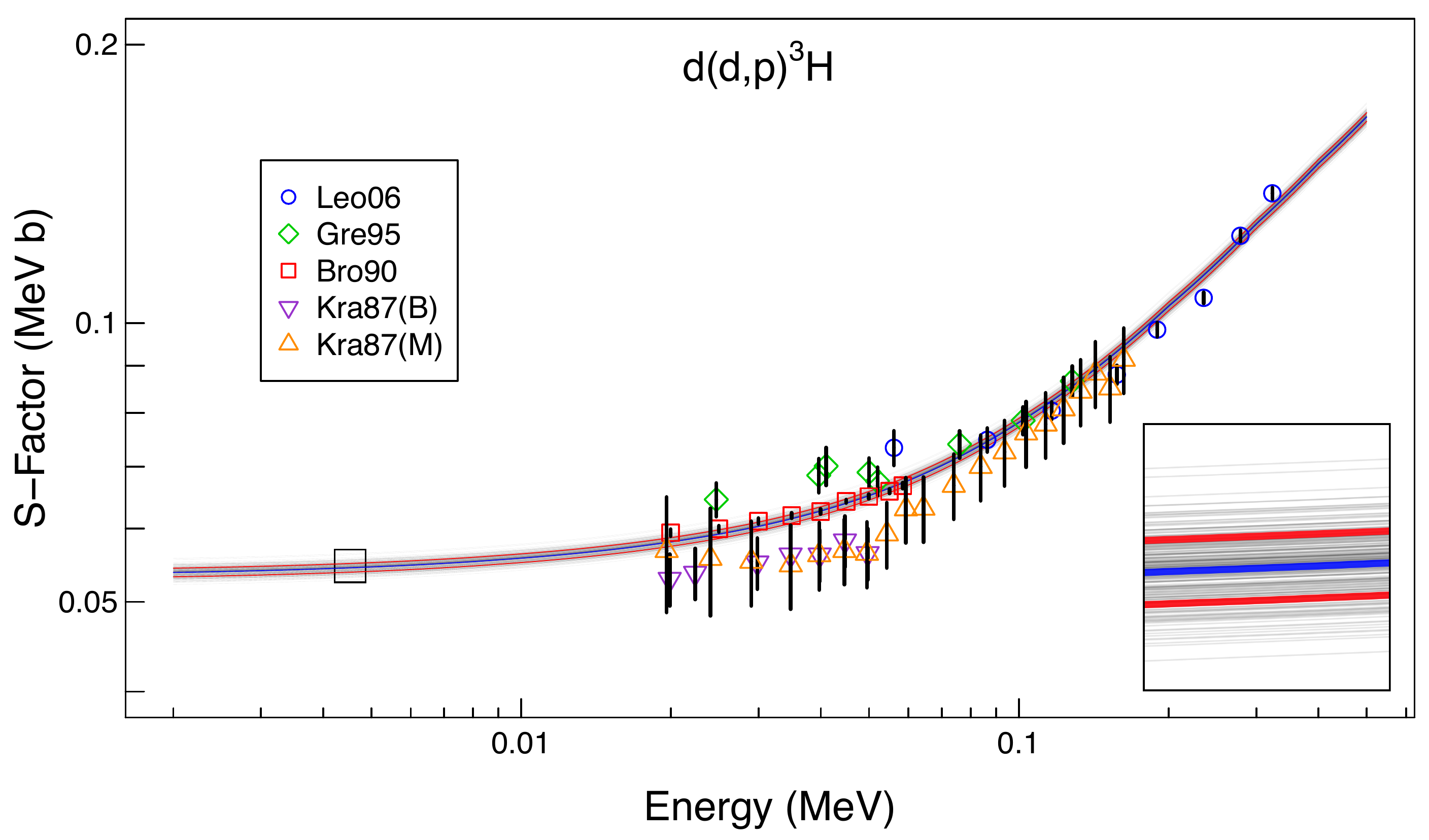}
			\caption{Astrophysical S-factor versus center-of-mass energy for the d(d,p)\textsuperscript{3}H reaction. The symbols show the data of \citet{leo} (circles), \citet{gre} (diamonds), \citet{bro} (squares), \citet{kra} (B) (down-pointing triangles) and \citet{kra} (M) (up-pointing triangles). The error bars (1$\sigma$) refer to statistical uncertainties only. Grey lines forming the shaded area correspond to credible S-factors that result from different sets of parameter samples (the inset shows a magnification for a clearer view of these lines). The blue line is the median (50th percentile) of all credible S-factors, and red lines correspond to the 16th and 84th percentiles. The credible lines are calculated from the theoretical S-factor of \citet{arai}, multiplied by a scale factor that is a parameter of the Bayesian model.}
			\label{fig:ResultsDdp}
\end{center}
\end{figure*}

\section{Strategy: Bayesian statistics and MCMC}\label{strategy}
We adopt the \textit{ab initio} calculation of \citet{arai} for the energy-dependence of the S-factor. This microscopic calculation uses a four-nucleon configuration space with a realistic nucleon-nucleon interaction. Their study was focused on low energies only, where partial waves up to J=2 contribute to the reaction cross section. 
Therefore, their calculation underestimates the data above a center-of-mass energy of 1 MeV. Consequently, we took only data points below an energy of 0.6 MeV into account in our Bayesian model.

We analyzed S-factor data by means of Bayesian statistics and Markov chain Monte Carlo (MCMC) algorithms. We used the software \texttt{JAGS} (``Just Another Gibbs Sampler") \citep{plummer}, specifically the \texttt{rjags} package, withing the R language \citep{rcore}. The inputs for the program are the experimental data \citep{bro,gre,kra,leo}\footnote{The experiments of \citet{kra} took place in M\"unster and at Bochum and so both data sets are considered independently.}, the theoretical nuclear model we want to scale, and the prior distributions of the model parameters (i.e., the scale factor of the theoretical S-factor curve and the normalization factors of each data set). 
The way of constructing the Markov chain in this project is by a Metropolis-Hastings algorithm. Each step of the chain consists in a set of values for all six parameters (the overall scale factor and the normalization factors of five data sets). The transition from one step to another can be summarized as:
\begin{enumerate}
\item Given a state $\theta^{(i)}$, propose a new one $\theta'$ by drawing a value from a proposal distribution (see \citet{albert}).
\item Accept the transition with a probability P$(\theta'|\theta^{(i)})$=min$(1,$$\frac{\text{P}(\theta'|S)}{\text{P}(\theta^{(i)}|S)})$, where $S$ stands for the experimental S-factor data. Moreover, P$(\theta|S)\propto$ P$(S|\theta)\cdot\pi(\theta)$, where P$(S|\theta)$ is the likelihood function and $\pi(\theta)$ is the prior distribution of the parameters. They are explained in Section \ref{sec:Likelihood and prior distributions}.
\item If the transition is accepted, $\theta^{(i+1)}=\theta'$. If not, $\theta^{(i+1)}=\theta^{(i)}$
\item Repeat 1-3.
\end{enumerate}

When the Markov chain reaches the steady state, the values of the parameters taken at every step yield their posterior distributions. With that information, lately it was possible to estimate the reaction rates. For more information about this method, see the Appendices in \citet{iliadis}. As a general reference in this topic, see \citet{hilbe}.

\subsection{Likelihood and prior distributions}
\label{sec:Likelihood and prior distributions}
The likelihood distribution of the S-factor given a set of parameters and prior distributions of those parameters are needed to compute the acceptance probabilities in the Markov chain.
The central limit theorem states that the probability density function resulting from the sum of independent random variables tends to a Gaussian distribution. 
By extension, a product of random variables will follow a lognormal distribution.
Measured nuclear reaction cross sections and astrophysical S-factors result from the product (or ratios) of different physical quantities. Thus we can assume that the likelihood function for the S-factor (P$(S|\theta)$ in Section \ref{strategy}) will follow a lognormal distribution \citep{longl}:

\begin{equation}
\label{eq:logn}
f(x)=\frac{1}{\sigma\sqrt{2\pi}x}e^{-(lnx-\mu)^2/(2\sigma^2)}, x>0
\end{equation}
\begin{center}
$\mu=\ln(E[x])-\dfrac{1}{2}\ln\bigg(1+\dfrac{V[x]}{E[x]^2}\bigg)$

$\sigma=\sqrt{\ln\bigg(1+\dfrac{V[x]}{E[x]^2}\bigg)}$
\end{center}
\noindent
where $\mu$ is the location parameter for the normally distributed logarithm of random variable $x$, i.e., $e^\mu$ is the median of the distribution of $x$, and $\sigma$ is the spread parameter for the normally distributed logarithm of $x$; E[x] and V[x] denote the expected mean value and the variance, respectively, of the lognormal distribution.
One advantage of this type of distribution is that negative S-factor values, which are unphysical, are not allowed. 
The lognormal likelihood function is then given by:
\begin{subequations}
	\begin{equation}
	\text{P}(\text{\textbf{S}}|\text{\textbf{\textit{f}}}) = \prod_{i=1}^{N} \frac{1}{S_i\sqrt{2\pi\sigma_{L;i}^2}} \text{ exp} \Bigg[ \frac{(\text{ln}\;S_i - \mu_i) ^2}{2\sigma_{L;i}^2} \Bigg]
	\end{equation}
	\begin{equation}
	\mu_i = \text{ln}\;(f_{n} f_{s} S_{th}) - \frac{1}{2} \text{ln}\; \big(1+\frac{\sigma_i^2}{(f_{n} f_{s} S_{th})^2}\big)
	\end{equation}
	\begin{equation}
	\sigma_{L;i}^2 = \text{ln} \big(1+\frac{\sigma_i^2}{(f_{n} f_{s} S_{th})^2}\big)
	\end{equation}
\end{subequations}
where $S_i$ stands for the experimental S-factor data, \textbf{\textit{f}} are the sampled parameters ($f_n$ is the normalization factor for a particular data set and $f_s$ is the overall scale factor), $N$ is the number of measurements of the data set, $\mu_i$ is the location parameter of data point $i$, $\sigma_{L;i}$ is the spread parameter of data point $i$, $S_{th}$ corresponds to the theoretical S-factor and $\sigma_i$ is the reported standard deviation of data point $i$.
Notice that there is no degeneracy regarding the product $f_n\cdot f_s$, since $f_n$ is different for each data set while $f_s$ is the same parameter throughout.

Since the scale factor, $f_s$, is expected to be close to unity, we assume for the overall scaling factor a non-informative prior ($\pi(\theta)$ in Section \ref{strategy}), i.e., a normally distributed probability density with a mean of zero and a standard deviation of 100. Therefore we expressed the prior for the scale factor as:
\begin{equation}
\pi(f_s) = 
	\begin{cases}
	\frac{1}{\sqrt{2\pi 100^2}} \text{exp}\Bigg[ \frac{(f_s-0.0)^2}{2\cdot100^2} \Bigg], & \text{for } f_s>0 \\
	0 & \text{for }f_s\leq0
	\end{cases}
\end{equation}

The distribution was truncated at zero since the scaling factor must be a positive quantity. To test the sensitivity of our results, we repeated the analysis using different priors (e.g., uniform distributions and gamma functions), and the results were very similar in all cases.
For the normalization factors of each data set, we assumed highly informative priors. It is discussed in Section \ref{sec:Systematic uncertainties}.

Additionally, we incorporate a robust regression method to avoid the bias that outliers can introduce in the results. Our algorithm accomplishes this by detecting possible outliers (i.e., measurements with over-optimistic uncertainties) and reducing their influence in the analysis (see Section \ref{robreg}).

\subsection{Systematic uncertainties}
\label{sec:Systematic uncertainties}
A measurement is usually subject to statistical and systematic uncertainties. Statistical uncertainties are inherent to any physical process and cannot be avoided. They can be reduced by combining results from different measurements, leading to different measured values for the same experimental conditions. 
Conversely, systematic uncertainties will not change if the experimental conditions remain the same. Hence, all of the data points from the same measurement will likely be affected by a systematic effect in a similar manner.
We introduce statistical uncertainties in our model by assuming lognormal priors for the individual normalization factors, $f_{n;k}$, of all five data sets, $k$.

The experimental data considered in this study \citep{bro,gre,kra,leo} provided systematic uncertainties for each data set as normalization factor uncertainties 1+$\epsilon$, with $\epsilon$ given in Table II of \citet{coc}. We include in our Bayesian model a systematic effect as a highly informative, lognormal prior. 
The parameters of this distribution are a median of 1.0, i.e., $e^\mu$ $=$ $1$, and a systematic factor uncertainty of $e^{\sigma_k}$. This prior can be written as:
\begin{equation}
\pi(f_{n;k}) = \frac{1}{f_{n;k} \sqrt{2\pi (\text{ln}(e^{\sigma_k}))^2}} \text{exp}\Bigg[ \frac{(\text{ln}\;f_{n;k} - \text{ln}(1.0)) ^2}{2(\text{ln}(e^{\sigma_k}))^2} \Bigg]
\end{equation}
For more information on this choice of prior, see \citet{iliadis}.

\subsection{Robust regression}
\label{robreg}
Outliers can bias the data analysis significantly and thus need to be treated carefully. In our \texttt{JAGS} code, we model outliers as data points with over-optimistic reported uncertainties. 
The algorithm designates each data point as either having believable uncertainty (i.e., not an outlier) or over-optimistic uncertainty (i.e., outlier). This operation is done for each step of the chain. Ultimately, data points having smaller outlier probabilities are more heavily weighted in the final results, thus reducing the statistical weight of the outliers \citep{robust}.
For the presentation of these results, we average the outlier probabilities for all data points in a given set and list the values in Tables \ref{tab:table1} and \ref{tab:table2}.

%

\section{Bayesian astrophysical S-factors}
The astrophysical S-factor of a nuclear reaction is defined as:
\begin{equation}
S(E)\equiv\sigma(E)Ee^{2\pi\eta}
\end{equation}
where $\sigma$(E) is the cross-section of the reaction at the center-of-mass energy E and $e^{2\pi\eta}$ is the Gamow factor, which depends on the charges of the projectile and the target, the relative atomic masses, and the energy E (see \citet{libro} for details).

The theoretical model used here for the energy dependence of the d+d S-factor is based on a multichannel \textit{ab initio} calculation \citep{arai}.
We assume that the nuclear model accurately predicts the energy dependence of the S-factor, but not necessarily its absolute scale. Our model predicts the best estimate of the overall scale factor and its uncertainty.

The Bayesian model for the analysis of the S-factor has several parameters. These include the normalization factors for each of the five individual data sets as well as the overall scale factor of the theoretical S-factor curve. 

We employ the same procedure as \citet{iliadis}, and we use three different Markov chains of 7500 steps each, with a burn-in of 2000 steps. These values ensure the convergence of the chains and that the Monte Carlo fluctuations are negligible compared to the statistical and systematic uncertainties. We performed several tests with different chain lengths (e.g., 75000 steps) and the results were the same.

\begin{figure}[ht]
\begin{center}
			\centering
			\includegraphics[width=8cm]{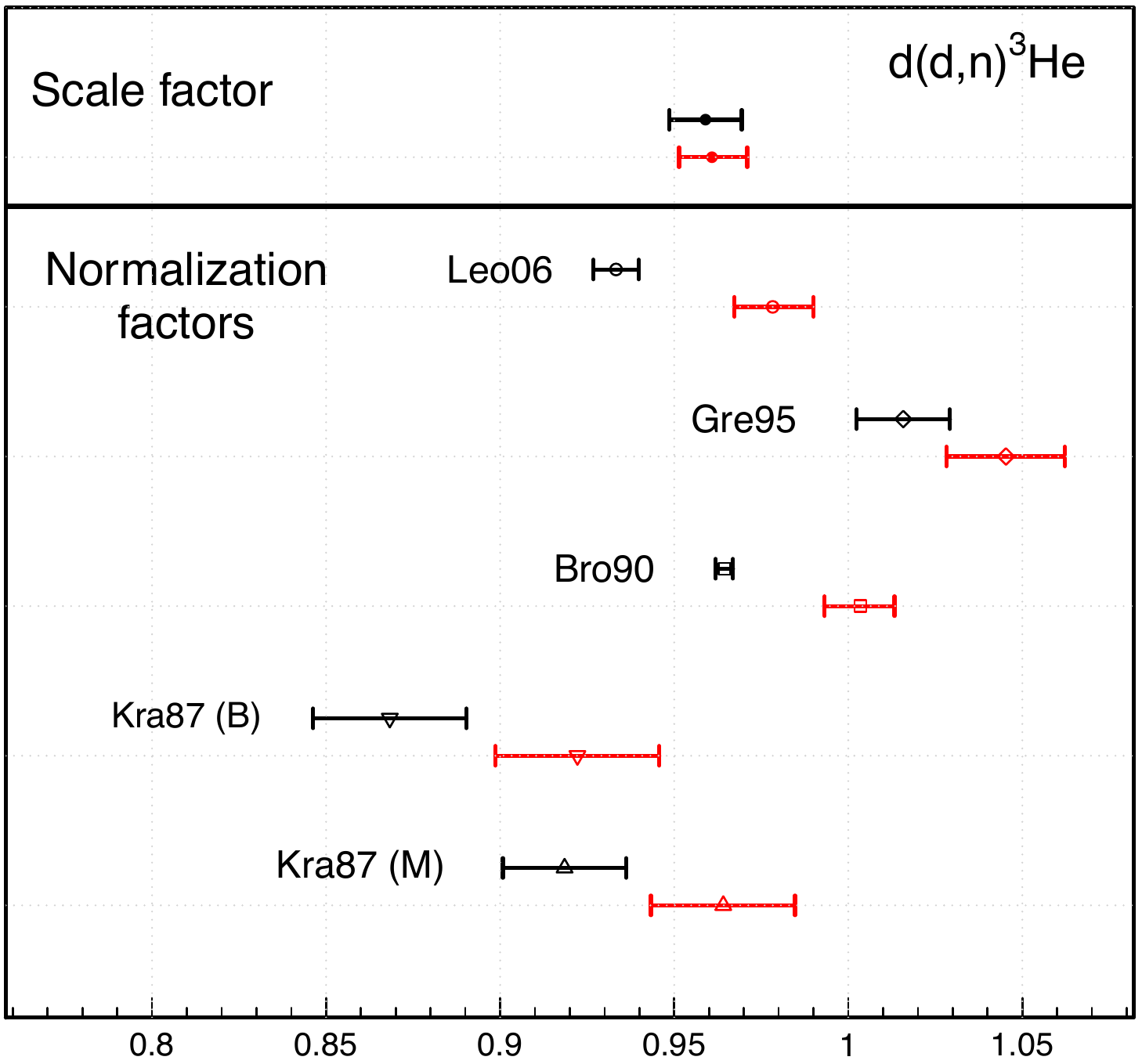}
			\caption{Results for d(d,n)\textsuperscript{3}He.
			(Top) Overall scale factor for the theoretical S-factor.
			(Bottom) Normalization factors of each data set: \citet{leo} (Leo06), \citet{gre} (Gre95), \citet{bro} (Bro90), \citet{kra} (Kra87 (B) and Kra87 (M)). 
			Present and previous \citep{coc} results are shown in red and black, respectively. The range indicated in red corresponds to the 68\% credible interval of the posterior. The range indicated in black shows the 68\% confidence interval of the traditional analysis.
			}
			\label{fig:comparison_ddn}
\end{center}
\end{figure}

\begin{figure}[hb]
\begin{center}
			\centering
			\includegraphics[width=8cm]{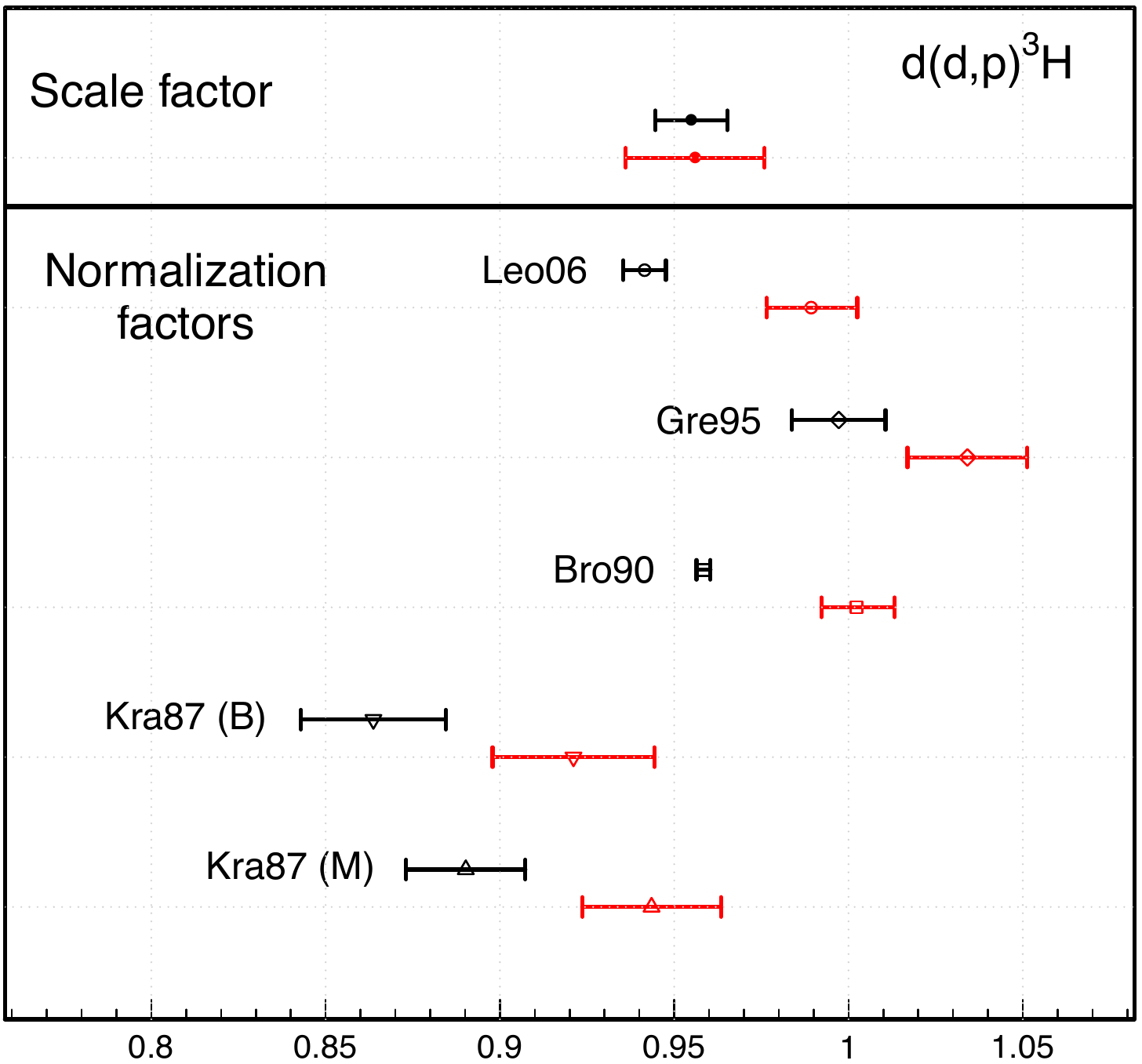}
			\caption{Results for d(d,p)\textsuperscript{3}H.
			(Top) Overall scale factor for the theoretical S-factor.
			(Bottom) Normalization factors of each data set: \citet{leo} (Leo06), \citet{gre} (Gre95), \citet{bro} (Bro90), \citet{kra} (Kra87 (B) and Kra87 (M)). 
			Present and previous \citep{coc} results are shown in red and black, respectively. The range indicated in red corresponds to the 68\% credible interval of the posterior. The range indicated in black shows the 68\% confidence interval of the traditional analysis.
			}
			\label{fig:comparison_ddp}
\end{center}
\end{figure}

\subsection{Results}

Traditional methods based on $\chi^2$ minimization have been applied to the calculation of the d+d reaction rates by \citet{coc}. In their analysis, they assumed that the scale factor is given by the weighted average of the normalization factors that independently fit each data set to the theoretical S(E) curve. 
They made a number of \textit{ad hoc} assumptions to include systematic errors in their analysis and assumed Gaussian approximations for the uncertainties (see Appendix A in \citet{coc}). Their results were deemed satisfactory by the authors, since the reduced $\chi^2$ was always close to unity.

Bayesian S-factors are shown in Figure \ref{fig:ResultsDdn} for d(d,n)\textsuperscript{3}He and Figure \ref{fig:ResultsDdp} for d(d,p)\textsuperscript{3}H. Grey lines represent credible S-factor curves for different sets of parameters, yielding the shaded region. All of the credible S-factors are close to the median value (blue line). The red lines correspond to the 16th and 84th percentiles.

Results from our Bayesian analysis, and the traditional method \citep{coc} for comparison, are shown in Tables \ref{tab:table1} and \ref{tab:table2} for the d(d,n)\textsuperscript{3}He and d(d,p)\textsuperscript{3}H reactions, respectively. Some of the results are also displayed in Figures \ref{fig:comparison_ddn} and \ref{fig:comparison_ddp}, where the red data points correspond to the present Bayesian method and the black data points correspond to the traditional $\chi^2$ minimization.
The top panels (labeled as ``Scale factor") display the overall scale factor. 
For both reactions, the scale factors are in agreement. It can also be seen that the scale factors are smaller than unity (see Tables \ref{tab:table1} and \ref{tab:table2}), i.e., the theoretical S-factor curve exceeds the data. The bottom regions (labeled as ``Normalization factors") of Figures \ref{fig:comparison_ddn} and \ref{fig:comparison_ddp} show the normalization factors of each data set. It can be seen that the Bayesian normalization factors are consistently larger than the traditional analysis values. This is caused by the different methods to calculate these factors, as explained below.

In the Bayesian approach, the theoretical S-factor is multiplied by the overall scale factor. We defined our Bayesian model so that each data set is the result of multiplying the scaled S-factor curve by a normalization factor. As explained before, this normalization factor includes the effect of systematic uncertainties. Hence, at each step of the Markov chain, there is a shift in the magnitude of the theory (scaling) and the data sets (to account for the systematic uncertainties). These shifts are performed by multiplying the S-factor theoretical curve by the overall scale factor and dividing each data set by its corresponding normalization factor. Each measurement is affected by a multiplicative error ($S_{exp}=f_n \cdot S_{true}$, where $S_{exp}$ is the experimental datum, $f_n$ is the normalization factor and $S_{true}$ is the actual value), so we must divide the experimental value by the normalization factor if we want to cancel it. In this way, the final probability density function for each parameter is influenced by all other parameters.

In the traditional analysis performed by \citet{coc}, however, the theoretical S-factor is multiplied by a normalization factor for each data set separately. The overall scale factor is then obtained by computing the weighted average of all normalization factors. The systematic uncertainties are introduced in the weights of the average by adding systematic and statistical errors quadratically for each data set (see Eq. (A8) in \citet{coc}).

\begin{figure}[ht]
\begin{center}
			\centering
			\includegraphics[width=8cm]{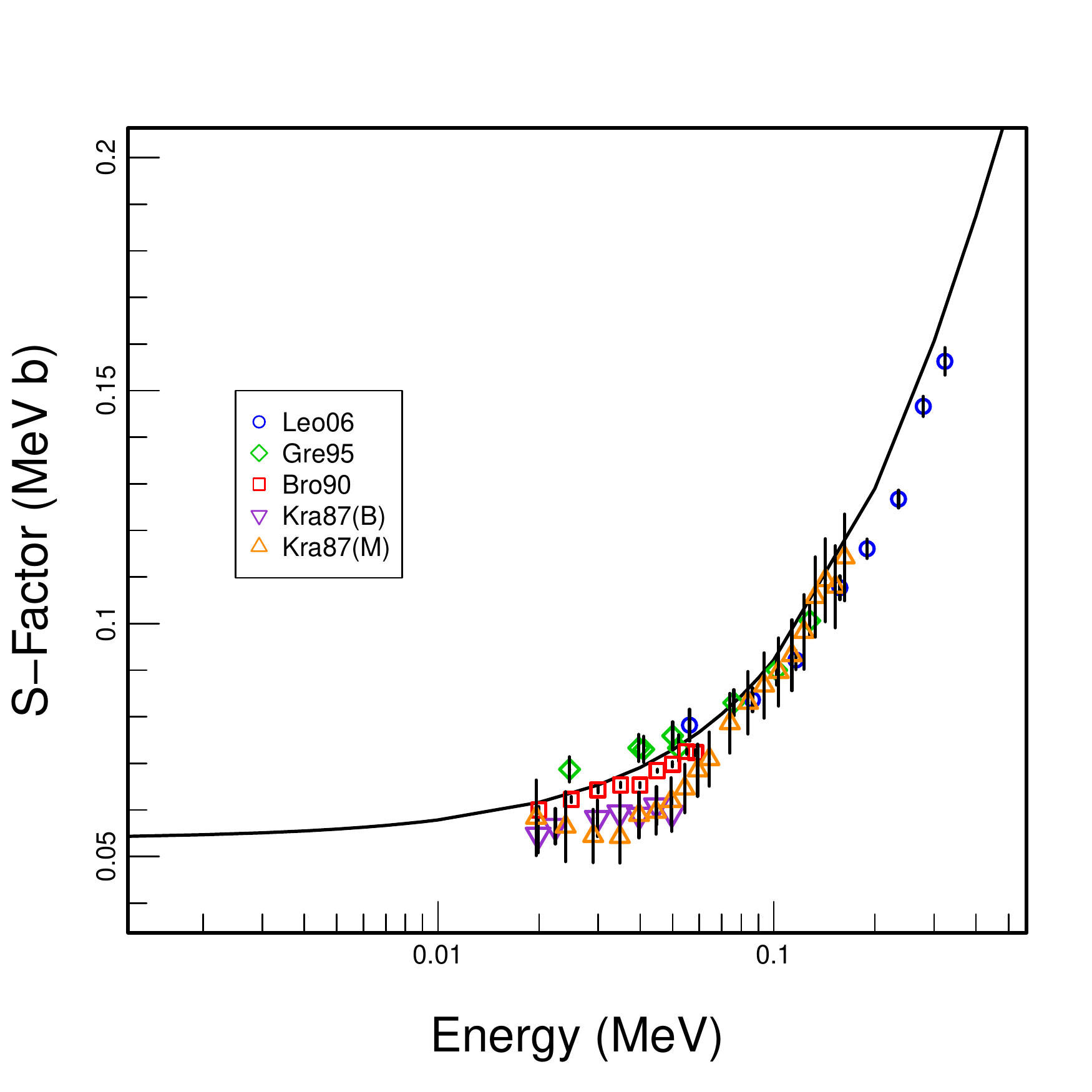}
			\caption{Astrophysical S-factor versus center-of-mass energy for d(d,n)\textsuperscript{3}He. Experimental points are from \citep{leo}, \citep{gre}, \citep{bro}, and \citep{kra}. The error bars (1$\sigma$) refer to statistical uncertainties only. The solid curve shows the \textit{ab initio} S-factor of \citet{arai} before scaling.}
			\label{fig:Prev_ddn}
\end{center}
\end{figure}

To explain the discrepancies between traditional and Bayesian normalization factors, consider the data presented in Figure \ref{fig:Prev_ddn}. This figure shows the measured d(d,n)$^3$He S-factors of each data set. The solid curve shows the ab initio S-factor of Arai et al. (2011) before scaling. At each step of the Markov chain, the Bayesian model suggests a new value smaller than unity for the overall scale factor, to displace the curve downwards. The model also samples a new normalization factor for each data set. As an example, look at the suggested normalization factor for Kra (B) in Table \ref{tab:table1} (0.922$\pm$0.024). It is less than unity since these experimental points should be shifted upwards to correct the effect of the systematic errors. Moreover, each normalization factor is influenced by the overall scale factor: at each step, the normalization factor fits the data to the scaled theory. In the traditional analysis, each normalization factor is calculated independently to fit the original S-factor. In the case of Kra (B), the traditional normalization factor needs to perform a larger shift, i.e. it will be further away from unity. This means a smaller normalization factor in the traditional case than in the Bayesian one.

\makeatletter
\newcommand\footnoteref[1]{\protected@xdef\@thefnmark{\ref{#1}}\@footnotemark}
\makeatother

\begin{table*}[ht]
\caption{Results for the d(d,n)\textsuperscript{3}He reaction.}
\label{tab:table1}
\begin{ruledtabular}
\begin{tabular}{cccccc} 
\multicolumn{2}{c}{Data} & \multicolumn{2}{c}{Present\footnote{\label{note1}Uncertainties derived from the 16th, 50th, and 84th percentiles.}} & \multicolumn{2}{c}{Previous\footnote{\label{note2}Data from \citet{coc}.}} \\
\cline{1-2}
\cline{3-4}
\cline{5-6}
  Ref.\footnote{Reference labels of data sets: Leo06 \citep{leo}, Gre95 \citep{gre}, Bro90 \citep{bro}, Kra87(B) \citep{kra}, and Kra87(M) \citep{kra}.}
   & n\footnote{Number of points of each data set.} & norm\footnote{Normalization factor for each data set (see explanation in text).} & outlier\footnote{Probability that the reported experimental uncertainty is over-optimistic. Calculated from average outlier probabilities of all data points in a given data set.} & norm\footnote{Normalization factor for each data set (see explanation in text). Uncertainties given represent $1\sigma$.} & $\chi^2_\nu$\footnote{Reduced $\chi^2$.}\\
  \hline
  Leo  06 & 8 & $0.978_{-0.011}^{+0.012}$ & 55.1\% & $0.933\pm0.007$ & 2.033\\
  Gre 95 & 8 & $1.045_{-0.017}^{+0.017}$ & 45.4\% & $1.016\pm0.013$ & 1.247\\
  Bro 90 & 9 & $1.004_{-0.010}^{+0.010}$ & 64.6\% & $0.964\pm0.003$ & 2.366\\
  Kra 87 (B) & 7 & $0.922_{-0.024}^{+0.024}$ & 35.3\% & $0.868\pm0.022$ & 0.292\\
  Kra 87 (M) & 20 & $0.964_{-0.021}^{+0.021}$ & 27.6\% & $0.919\pm0.018$ & 0.624\\
  \\
  \hline
  \\
  \multicolumn{2}{c}{Quantity} & \multicolumn{2}{c}{Present\footnoteref{note1}} & \multicolumn{2}{c}{Previous\footnoteref{note2}} \\
\cline{1-2}
\cline{3-4}
\cline{5-6}
  \hline
    \multicolumn{2}{c}{Scale factor\footnote{Best estimate for the scale factor of the theoretical S-factor from \citet{arai}.}:} & \multicolumn{2}{c}{$0.961_{-0.010}^{+0.010}$} & \multicolumn{2}{c}{$0.959\pm0.010$ $(\chi^2_\nu=1.33)$} \\
    \multicolumn{2}{c}{S(0) (keVb):} & \multicolumn{2}{c}{$51.70_{-0.51}^{+0.54}$} & \multicolumn{2}{c}{-} \\
  \hline
\end{tabular}
\end{ruledtabular}
\end{table*}
\begin{table*}[ht]
\caption{Results for the d(d,p)\textsuperscript{3}H reaction.}
\label{tab:table2}
\begin{ruledtabular}
\begin{tabular}{cccccc} 
\multicolumn{2}{c}{Data} & \multicolumn{2}{c}{Present\footnote{\label{note3}Uncertainties derived from the 16th, 50th, and 84th percentiles.}} & \multicolumn{2}{c}{Previous\footnote{\label{note4}Data from \citet{coc}.}} \\
\cline{1-2}
\cline{3-4}
\cline{5-6}
  Ref.\footnote{Reference labels of data sets: Leo06 \citep{leo}, Gre95 \citep{gre}, Bro90 \citep{bro}, Kra87(B) \citep{kra}, and Kra87(M) \citep{kra}.}
   & n\footnote{Number of points of each data set.} & norm\footnote{Normalization factor for each data set (see explanation in text).} & outlier\footnote{Probability that the reported experimental uncertainty is over-optimistic. Calculated from average outlier probabilities of all data points in a given data set.} & norm\footnote{Normalization factor for each data set (see explanation in text). Uncertainties given represent $1\sigma$.} & $\chi^2_\nu$\footnote{Reduced $\chi^2$.}\\
  \hline
  Leo  06 & 8 & $0.989_{-0.013}^{+0.013}$ & 80.4\% & $0.942\pm0.006$ & 5.376\\
  Gre 95 & 8 & $1.034_{-0.017}^{+0.017}$ & 30.6\% & $0.997\pm0.013$ & 0.999\\
  Bro 90 & 9 & $1.002_{-0.010}^{+0.011}$ & 51.9\% & $0.958\pm0.002$ & 1.969\\
  Kra 87 (B) & 7 & $0.921_{-0.023}^{+0.023}$ & 21.1\% & $0.864\pm0.021$ & 0.100\\
  Kra 87 (M) & 20 & $0.944_{-0.020}^{+0.020}$ & 8.6\% & $0.890\pm0.017$ & 0.177\\
  \\
  \hline
  \\
  \multicolumn{2}{c}{Quantity} & \multicolumn{2}{c}{Present\footnoteref{note3}} & \multicolumn{2}{c}{Previous\footnoteref{note4}} \\\cline{1-2}
\cline{3-4}
\cline{5-6}
  \hline
    \multicolumn{2}{c}{Scale factor\footnote{Best estimate for the scale factor of the theoretical S-factor from \citet{arai}.}:} & \multicolumn{2}{c}{$0.956_{-0.011}^{+0.010}$} & \multicolumn{2}{c}{$0.955\pm0.010$ $(\chi^2_\nu=1.33)$} \\
    \multicolumn{2}{c}{S(0) (keVb):} & \multicolumn{2}{c}{$53.26_{-0.59}^{+0.55}$} & \multicolumn{2}{c}{-} \\
  \hline
\end{tabular}
\end{ruledtabular}
\end{table*}
%
\clearpage
\section{Reaction rates}
\begin{figure*}[ht]
\begin{center}
			\centering
			\includegraphics[width=\textwidth]{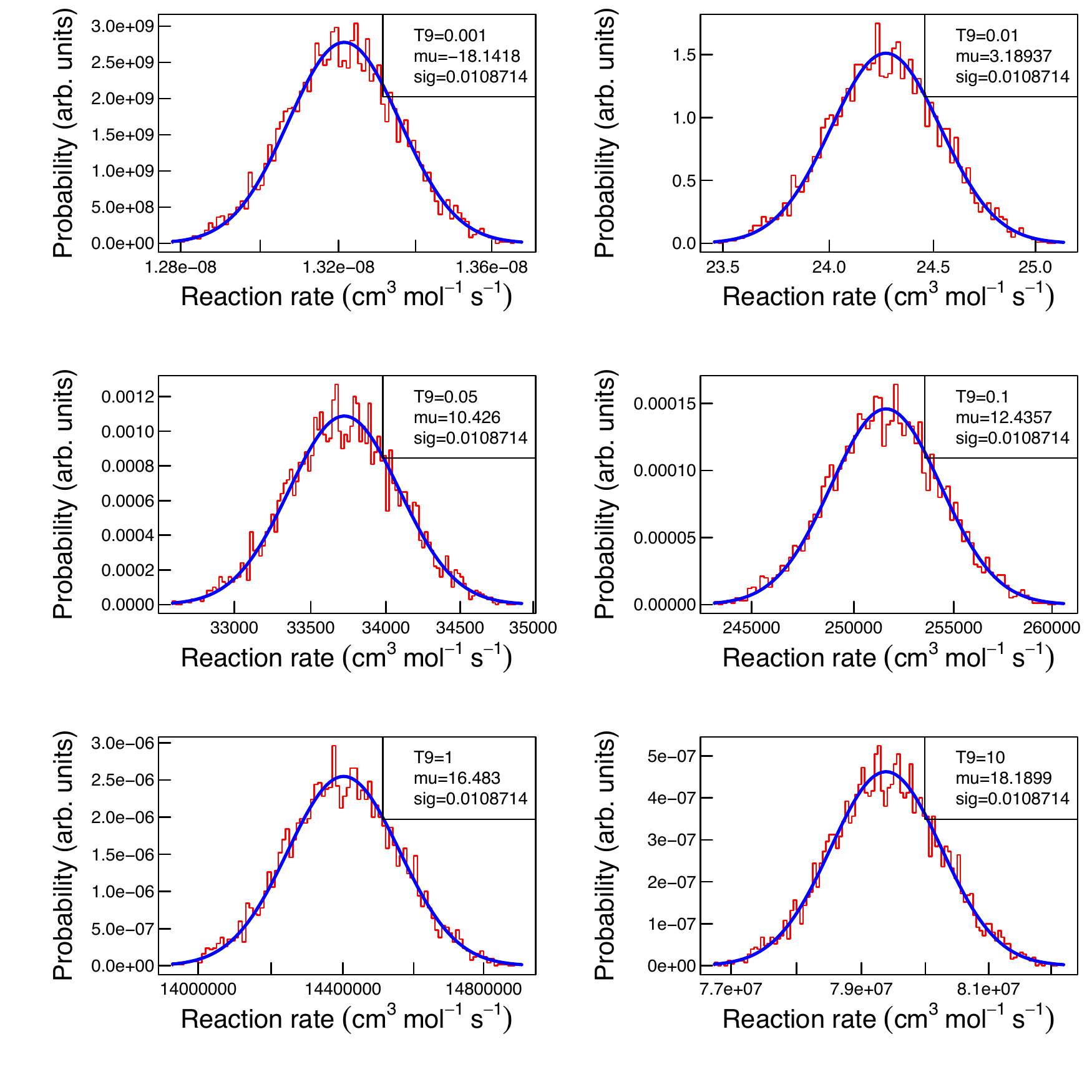}
			\caption{Reaction rate probability density of d(d,n)\textsuperscript{3}He at different temperatures. The rate samples (red histograms) are computed using the S-factor samples obtained from the Bayesian analysis. Blue curves represent lognormal approximations, where the lognormal parameters $\mu$ (`mu") and $\sigma$ (``sig") are directly calculated from the expectation value and variance of all rate samples, ln($N_A\left\langle\sigma v\right\rangle_i$), at a given temperature. T9 is the temperature in GK.}
			\label{fig:ratesddn}
\end{center}
\end{figure*}

\begin{figure*}[ht]
\begin{center}
			\centering
			\includegraphics[width=\textwidth]{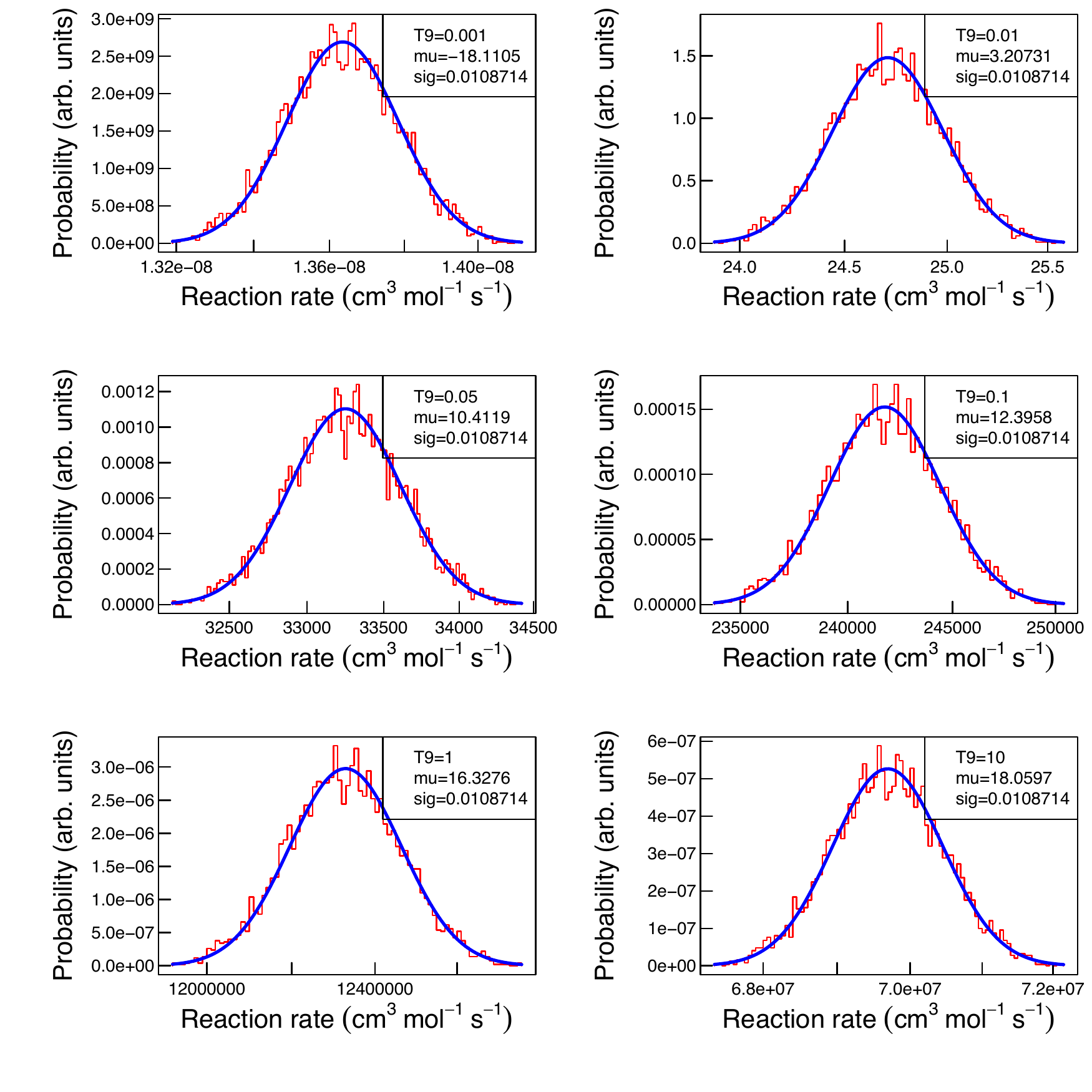}
			\caption{Reaction rate probability density of d(d,p)\textsuperscript{3}H at different temperatures. The rate samples (red histograms) are computed using the S-factor samples obtained from the Bayesian analysis. Blue curves represent lognormal approximations, where the lognormal parameters $\mu$ (`mu") and $\sigma$ (``sig") are directly calculated from the expectation value and variance of all rate samples, ln($N_A\left\langle\sigma v\right\rangle_i$), at a given temperature. T9 is the temperature in GK.}
			\label{fig:ratesddp}
\end{center}
\end{figure*}

The thermonuclear reaction rate per particle pair, $N_A\left\langle\sigma v\right\rangle$, can be written as:
\begin{equation}
	\label{eq:rate}
	\begin{split}
	N_A\left\langle\sigma v\right\rangle=&\bigg(\frac{8}{\pi m_{01}}\bigg)^{1/2}\frac{N_A}{(kT)^{3/2}}\\
	&\int_0^\infty e^{-2\pi \eta}S(E)e^{-E/kT}dE
	\end{split}
\end{equation}
where $m_{01}$ is the reduced mass of projectile and target, $N_A$ represents Avogadro's constant, and the product of Boltzmann constant, $k$, and plasma temperature, $T$, is given by
\begin{equation}
	\label{eq:kT}
	kT=0.086173324 \: T_9\text{ (MeV)}
\end{equation}
with the temperature, $T_9$, in units of GK (see \citet{libro} for details).
The reaction rates are calculated by numerical integration of Eq. (\ref{eq:rate}) for each set of parameters sampled by the Markov chain, at 60 different temperatures between 1 MK and 10 GK. The reaction rate probability densities at selected temperatures are shown in Figures \ref{fig:ratesddn} and \ref{fig:ratesddp} in red. The blue lines correspond to a lognormal approximation \citep{longl}, for convenient implementation of the rates in libraries such as STARLIB \citep{starlib}. 
Numerical reaction rate values are listed in Table \ref{tab:table3}. The recommended rates are computed as the 50th percentile of the probability density, while the rate factor uncertainty, f.u., is obtained from the 16th and 84th percentiles. The lognormal parameters, $\mu$ and $\sigma$, can be calculated from the recommended (median) rate ($x_{med} = e^\mu$) and the factor uncertainty ($f.u.=e^\sigma$; for a coverage probability of 68\%).
The rate factor uncertainty is 1.1\% for both reactions at most temperatures.

The present rates for d(d,n)\textsuperscript{3}He and d(d,p)\textsuperscript{3}H agree with the results of \citet{coc} within 1\% at most temperatures. However, our rates are more than 15\% larger than those of \citet{coc} at very low temperatures (near 1 MK). 
This is caused by a low-energy cutoff that is too high for the numerical integration of the rates in the previous analysis. 
The theoretical model of \citet{arai} only applies to low energies, and thus we can derive Bayesian reaction rates only up to a temperature of 2 GK. The results in Table \ref{tab:table3} for higher temperatures, shown in italics, are adopted from \citet{coc}. 
The most important temperatures for BBN are near 1 GK, corresponding to an effective kinetic energy range of $<$250 keV for the d+d reactions.

The last step is to calculate the effect of the new reaction rates on the predicted primordial D/H ratio. The Bayesian mean value for the scale factor is larger by 0.21\% for d(d,n)\textsuperscript{3}He and larger by 0.12\% for d(d,p)\textsuperscript{3}H compared to \citet{coc}. The discrepancies of both reaction rates (0.21\% and 0.12\%, respectively), weighted by the sensitivity of the D/H abundance ratio to each reaction rate variation (-0.54 and -0.46, respectively) \citep{coco}, result in a 0.113\% and 0.055\% decrease of the central D/H value. 
Fortuitously, the uncertainties on the scale factors (see Tables \ref{tab:table1} and \ref{tab:table2}) are almost identical to the former ones \citep{coc}. Hence, when using these two new reaction rates, instead of the \citet{coc} ones, this translates to a 0.16\% decrease of the predicted D/H value, while its total uncertainty remains unchanged at 2.0\%. Half of this error budget originates from the d(p,$\gamma$) reaction rate and it would be premature 
to update the D/H value before new measurements concerning this reaction, done at LUNA, are published (see \citet{mossa}). Only after these new data are made available
and the investigation of other sources of uncertainties (numerical, correction to weak rates,...) are completed, it will be relevant to provide new predictions of D/H.


\onecolumngrid
\begin{deluxetable}{ccccc}
\tablewidth{0pt}
\tablecaption{Present recommended reaction rates.\tablenotemark{a}\label{tab:table3}}
\tablehead{
 & \multicolumn{2}{c}{d(d,n)\textsuperscript{3}He}  & \multicolumn{2}{c}{d(d,p)\textsuperscript{3}H} \\
\cline{2-5}
\colhead{T (GK)} & \colhead{Rate} & \colhead{$f.u.$} &   \colhead{Rate} & \colhead{$f.u.$} 
}
\startdata
0.001	&	1.322E-08	&	1.011	&	1.364E-08	&	1.011	\\	
0.002	&	5.489E-05	&	1.011	&	5.653E-05	&	1.011	\\	
0.003	&	3.025E-03	&	1.011	&	3.110E-03	&	1.011	\\	
0.004	&	3.737E-02	&	1.011	&	3.835E-02	&	1.011	\\	
0.005	&	2.214E-01	&	1.011	&	2.269E-01	&	1.011	\\	
0.006	&	8.556E-01	&	1.011	&	8.755E-01	&	1.011	\\	
0.007	&	2.508E+00	&	1.011	&	2.563E+00	&	1.011	\\	
0.008	&	6.074E+00	&	1.011	&	6.198E+00	&	1.011	\\	
0.009	&	1.280E+01	&	1.011	&	1.304E+01	&	1.011	\\	
0.010	&	2.427E+01	&	1.011	&	2.471E+01	&	1.011	\\	
0.011	&	4.242E+01	&	1.011	&	4.314E+01	&	1.011	\\	
0.012	&	6.945E+01	&	1.011	&	7.055E+01	&	1.011	\\	
0.013	&	1.078E+02	&	1.011	&	1.094E+02	&	1.011	\\	
0.014	&	1.602E+02	&	1.011	&	1.624E+02	&	1.011	\\	
0.015	&	2.293E+02	&	1.011	&	2.322E+02	&	1.011	\\	
0.016	&	3.183E+02	&	1.011	&	3.220E+02	&	1.011	\\	
0.018	&	5.674E+02	&	1.011	&	5.729E+02	&	1.011	\\	
0.020	&	9.321E+02	&	1.011	&	9.395E+02	&	1.011	\\	
0.025	&	2.507E+03	&	1.011	&	2.516E+03	&	1.011	\\	
0.030	&	5.307E+03	&	1.011	&	5.305E+03	&	1.011	\\	
0.040	&	1.570E+04	&	1.011	&	1.558E+04	&	1.011	\\	
0.050	&	3.373E+04	&	1.011	&	3.325E+04	&	1.011	\\	
0.060	&	6.020E+04	&	1.011	&	5.900E+04	&	1.011	\\	
0.070	&	9.539E+04	&	1.011	&	9.298E+04	&	1.011	\\	
0.080	&	1.392E+05	&	1.011	&	1.350E+05	&	1.011	\\	
0.090	&	1.914E+05	&	1.011	&	1.847E+05	&	1.011	\\	
0.100	&	2.516E+05	&	1.011	&	2.418E+05	&	1.011	\\	
0.110	&	3.194E+05	&	1.011	&	3.056E+05	&	1.011	\\	
0.120	&	3.943E+05	&	1.011	&	3.758E+05	&	1.011	\\	
0.130	&	4.759E+05	&	1.011	&	4.518E+05	&	1.011	\\	
0.140	&	5.638E+05	&	1.011	&	5.334E+05	&	1.011	\\	
0.150	&	6.575E+05	&	1.011	&	6.199E+05	&	1.011	\\	
0.160	&	7.568E+05	&	1.011	&	7.111E+05	&	1.011	\\	
0.180	&	9.702E+05	&	1.011	&	9.061E+05	&	1.011	\\	
0.200	&	1.201E+06	&	1.011	&	1.116E+06	&	1.011	\\	
0.250	&	1.843E+06	&	1.011	&	1.691E+06	&	1.011	\\	
0.300	&	2.555E+06	&	1.011	&	2.321E+06	&	1.011	\\	
0.350	&	3.318E+06	&	1.011	&	2.988E+06	&	1.011	\\	
0.400	&	4.118E+06	&	1.011	&	3.681E+06	&	1.011	\\	
0.450	&	4.944E+06	&	1.011	&	4.391E+06	&	1.011	\\	
0.500	&	5.788E+06	&	1.011	&	5.113E+06	&	1.011	\\	
0.600	&	7.510E+06	&	1.011	&	6.573E+06	&	1.011	\\	
0.700	&	9.251E+06	&	1.011	&	8.036E+06	&	1.011	\\	
0.800	&	1.099E+07	&	1.011	&	9.489E+06	&	1.011	\\	
0.900	&	1.271E+07	&	1.011	&	1.092E+07	&	1.011	\\	
1.000	&	1.440E+07	&	1.011	&	1.233E+07	&	1.011	\\	
1.250	&	1.850E+07	&	1.011	&	1.572E+07	&	1.011	\\	
1.500	&	2.236E+07	&	1.011	&	1.893E+07	&	1.011	\\	
1.750	&	2.599E+07	&	1.011	&	2.194E+07	&	1.011	\\	
2.000	&	2.938E+07	&	1.011	&	2.477E+07	&	1.011	\\	
2.500	&	\textit{3.546E+07}	&	\textit{1.012}	&	\textit{2.976E+07}	&	\textit{1.013}	\\	
3.000	&	\textit{4.093E+07}	&	\textit{1.014}	&	\textit{3.440E+07}	&	\textit{1.014}	\\	
3.500	&	\textit{4.585E+07}	&	\textit{1.014}	&	\textit{3.863E+07}	&	\textit{1.014}	\\	
4.000	&	\textit{5.031E+07}	&	\textit{1.015}	&	\textit{4.251E+07}	&	\textit{1.015}	\\	
5.000	&	\textit{5.816E+07}	&	\textit{1.016}	&	\textit{4.946E+07}	&	\textit{1.016}	\\	
6.000	&	\textit{6.488E+07}	&	\textit{1.017}	&	\textit{5.552E+07}	&	\textit{1.017}	\\	
7.000	&	\textit{7.072E+07}	&	\textit{1.018}	&	\textit{6.077E+07}	&	\textit{1.018}	\\	
8.000	&	\textit{7.583E+07}	&	\textit{1.018}	&	\textit{6.529E+07}	&	\textit{1.018}	\\	
9.000	&	\textit{8.037E+07}	&	\textit{1.018}	&	\textit{6.912E+07}	&	\textit{1.018}	\\	
10.000	&	\textit{8.437E+07}	&	\textit{1.018}	&	\textit{7.228E+07}	&	\textit{1.019}	\\	
\enddata
\tablenotemark{a}{Reaction rates in units of cm\textsuperscript{3}mol\textsuperscript{-1}s\textsuperscript{-1}, corresponding to the 50th percentile of the rate probability density function. The rate factor uncertainty, f.u., is obtained from the 16th and 84th percentiles (see text). The parameters $\mu$ and $\sigma$ of the lognormal approximation to the reaction rate are given by $x_{med} = e^\mu$ and $f.u. = e^\sigma$, respectively, where $x_{med}$ denotes the median rate. Values for $T>2$ GK, shown in italics, are adopted from \citet{coc}.}
\end{deluxetable}

\twocolumngrid


\section{Conclusions}
We presented improved reaction rates for d(d,n)\textsuperscript{3}He and d(d,p)\textsuperscript{3}H based on the Bayesian method discussed in \citet{iliadis}.
Unlike previous methods that were based on traditional statistics (i.e., $\chi^2$ minimization), our method does not rely on weighted averages or the quadratic addition of systematic and statistical errors.
For both reactions, the rate factor uncertainty is 1.1\% and agrees with the traditional results. However, the Bayesian scale factors by which the theory needs to be multiplied to fit the data are larger than those of \citet{coc}. We obtained scale factors which are 0.20\% larger for d(d,n)\textsuperscript{3}He and 0.12\% larger for d(d,p)\textsuperscript{3}H. This translates to a 0.16\% decrease of the predicted D/H value, while its total uncertainty remains unchanged at 2.0\%. This shows the robustness of the deuterium predictions, provided that the same experimental data and nuclear model are used. It leaves very little room for those solutions to the lithium problem that cannot avoid an increase in D/H. It also calls for improved theoretical calculations.
The theoretical work of \citet{arai}, used here,  was focused on low energies and does not correctly reproduce the experimental data above $\approx$600 keV. 
It is highly desirable that these calculations be extended up to $\approx$2 MeV, to cover the range of experimental data.

Here we presented the first statistically rigorous results for d+d reaction rate probability densities. These can be employed in future Monte Carlo studies of big bang nucleosynthesis.


\section{Acknowledgements}
We would like to thank Jordi Jos\'{e}, Jack Dermigny, Rafa De Souza, Lori Downen and Sean Hunt for their support and feedback. One of us (AGI) would like to express his gratitude to the Department of Physics and Astronomy for hospitality during his visit to UNC-CH, where this project was started.
This work was supported in part by NASA under the Astrophysics Theory Program grant 14-ATP14-0007 and the U.S. DOE under Contract No. DE-FG02-97ER41041.

\software{JAGS \citep{plummer}, R \citep{rcore}}

\end{document}